\documentclass[12pt]{iopart}
\usepackage[utf8]{inputenc}
\usepackage[square,numbers]{natbib}
\usepackage{graphicx}
\usepackage{color}
\usepackage{bm}
\usepackage{amssymb}
\usepackage[normalem]{ulem}
\usepackage[dvipsnames]{xcolor}
\usepackage[colorlinks,urlcolor=NavyBlue,citecolor=NavyBlue,linkcolor=NavyBlue,pdfusetitle]{hyperref}

\newcommand{\ud}{\mathrm{d}}

\newcommand{\msun}{\,\mathrm{M}_\odot}

\begin{document}

\title[Supernova GW Signals: Theory and Modeling]{Core-Collapse Supernovae and their Gravitational Wave Signals: The Status of Theory and Modeling}
\author{Bernhard M\"uller$^{1}$}
\address{$^{1}$School of Physics and
 Astronomy, Monash University, Clayton, VIC 3800, Australia}
\ead{bernhard.mueller@monash.edu}
\vspace{10pt}
\begin{indented}
\item[]\today
\end{indented}

\begin{abstract}
The detection of gravitational waves from a core-collapse supernova in the Milky Way or its vicinity represents a unique opportunity to probe the inner workings of these explosions. In this review, I briefly summarize our current understanding of the supernova explosion mechanism and then outline the physical processes that shape the supernova gravitational wave signal. The review highlights how the various components of the signal have the potential to constrain the progenitor rotation, the proto-neutron star structure, the nuclear equation of state, the nature of hydrodynamic instabilities, and the violence of turbulent motions in the supernova core. I also highlight some open questions and uncertainties in the theory of supernova gravitational wave astronomy as well as challenges for further progress. Specifically, there is a need to develop large model databases, systematic uncertainty quantification and methods for evidence assessment to prepare for multi-messenger observations from a Galactic supernova.
\end{abstract}

\section{Introduction}
\label{sec:intro}
Core-collapse supernovae, the explosions of massive stars, are perhaps the most precious target for multi-messenger astronomy. Historically, the detection of about two dozen neutrinos from SN~1987A \citep{hirata_87,bionta_87,alekseev_87} inaugurated extragalactic multi-messenger astronomy with the first observation of an astrophysical transient both inside and outside the electromagnetic spectrum. Core-collapse supernovae have long received considerable interest as a source of gravitational waves (GWs) from the early days of theoretical gravitational wave astrophysics
\citep{press_72}. Advances in simulations have brought the predictions of GW amplitudes from supernovae down since these early days, limiting the anticipated detection horizon to the Milky Way and, in optimistic scenarios, its nearby extragalactic environment, and reducing the expected GW event rate accordingly. Nonetheless, a  Galactic supernova undoubtedly constitutes the biggest opportunity for multi-messenger astronomy as yet with the first concurrent detection of electromagnetic emission, neutrinos, and gravitational waves. Different from kilonovae from compact binary mergers \citep{abbott_17}, a galactic supernova will leave a permanent ``monument'' -- the transient and later the supernova remnant -- providing data for further study indefinitely unless the explosion is obscured by dust. In that latter case, gravitational waves and neutrinos would become indispensable messengers for signaling the collapse of a massive star and its outcome.

Based on two lectures at the recent event
\emph{SN2025gw: First IGWN Symposium on Core Collapse Supernova Gravitational Wave Theory and Detection}, this review attempts to concisely summarize the current state of the theory 
of GW emission from core-collapse supernovae. There are a number of reviews and original papers that summarize much of the  physics of GW emission \citep[e.g.,][]{yakunin_10,mueller_13,radice_19,abdikamalov_22,mezzacappa_25}, and I therefore do not attempt to replicate all of these efforts. 

Emphasis is placed on the key physical dependencies, principal uncertainties, and unresolved questions. The state of research on the core-collapse supernova explosion mechanism and simulation methodologies are only addressed in a cursory and qualitative manner in Section~\ref{sec:mechanism} to the extent that it is required as a basis for gravitational wave astronomy. I refer to other recent reviews on these subjects \citep{mueller_20,mezzacappa_20,burrows_21,janka_25} for in-depth information. This is followed by an overview of the theory of GW emission and current results from supernova simulations in Section~\ref{sec:gw}.
In the light of discussions at the symposium, I will also outline challenges in preparing for a galactic supernova as a rare high-reward event, and possible strategies for supernova GW astronomy to address these in Section~\ref{sec:conclusions}.

\section{The Dynamics of Collapse and Explosion}
\label{sec:mechanism}
\subsection{The Progenitors of Core-Collapse Supernovae}
In massive stars, stellar evolution proceeds 
quasi-hydrostatically through H, He, C, Ne, O, and Si burning (in core and shell burning episodes) up to the formation of an iron core, leaving a characteristic ``onion-shell'' structure \citep{woosley_02}. For single stars, this evolutionary pathway covers the mass range between about $8\msun$ and $130\msun$ on the zero-age main sequence \citep{heger_03}. The precise limits depend on metallicity, and a large fraction of massive stars experience binary mass transfer \citep{sana_12}, which affects the final fate as well. The demarcation line at the low-mass end separates massive stars from super asymptotic giant branch (Super-AGB) stars \citep{nomoto_84,jones_14,doherty_17,leung_19a,woosley_15}, which ignite carbon, but experience degeneracy effects during and after carbon burning. The more massive Super-AGB stars may share a similar fate as massive stars as ``electron-capture supernovae'' 
\citep{nomoto_84,nomoto_87} (see below). At the high-mass end, there is a transition to the pair instability regime \citep{heger_03,woosley_17} where collapse is triggered by electron-positron pair formation after C  burning. For ZAMS masses from about $85\msun$ to $130\msun$ for single stars \citep{heger_03,woosley_17,heger_23} (with mass limits subject to uncertainties), this instability results in partial mass ejection by pulses (pulsational pair instability supernova), but the star eventually evolves further up to iron core formation as a massive star. At higher masses, complete disruption
by oxygen burning occurs (pair instability supernova), and above $\mathord{\sim} 260\,\msun$ pair instability leads to direct black hole formation.

\subsection{Collapse and Bounce}
\label{sec:bounce}
For massive stars, core collapse is eventually triggered when the iron core reaches a sufficiently high mass after silicon core and shell burning. As the core approaches its effective Chandrasekhar mass, it reaches sufficiently high densities to enable electron captures on heavy nuclei and to some extent on free protons \citep{bethe_90,langanke_03}, which reduce the electron degeneracy pressure in the core. Further contraction increases the rates of electron capture, leading to  gravitational collapse on a free-fall time scale. Especially for progenitors of higher mass with less degenerate cores, photodissociation of heavy nuclei also contributes to the initiation of collapse.
In the aforementioned electron-capture supernovae from Super-AGB stars, dynamical collapse is already triggered after the formation of an oxygen-neon-magnesium core by
electron captures on $^{20}\mathrm{Ne}$
and $^{24}\mathrm{Mg}$. The width of the electron-capture
channel remains subject to further research
due to the complex interplay of neutrino losses,
nuclear burning and convection in this transition regime between intermediate-mass and high-mass stars \citep{jones_14,jones_16,doherty_17,kirsebom_19}. Similarly complex pre-collapse dynamics can occur in the least massive supernovae progenitors with iron cores \citep{woosley_15}.

Electron capture removes lepton number from the core until densities of  $\mathord{\sim} 10^{12}\,\mathrm{g}\,\mathrm{cm}^{-3}$ are reached. At this point, the emitted electron neutrinos are trapped, and the lepton fraction remains constant for the remainder of the collapse phase. Collapse is stopped after the core overshoots nuclear density and ``bounces'' due to the stiffening of the equation of state (EoS). The rebound launches a shock wave, whose initial kinetic energy of $\sim 10^{51}\mathrm{erg}$ is quickly drained by neutrino losses and the dissociation of heavy nuclei in the shocked matter \citep{mazurek_82,burrows_85}. The shock ``stalls'' within milliseconds after bounce, i.e., it turns into an accretion shock. 
As the outer shells of the core are accreted
onto what is now a ``proto-neutron star'' and accretion rates drop, the accretion shock still moves out to a radius of about $150\,\mathrm{km}$ before starting to recede again. The supernova explosion is triggered later by some mechanism that injects energy into the post-shock region (Sections~\ref{sec:sh_revival}--\ref{sec:pt}).

Once the  accretion shock moves to densities where neutrinos are no longer trapped, rapid deleptonization of the shocked shells occurs and produces an electron neutrino
burst with peak luminosities of $\mathord{\sim}4\times 10^{53}\,\mathrm{erg}\,\mathrm{s}^{-1}$. The neutrino losses as well as the decreasing shock strength then produce a negative entropy gradient behind the shock, which triggers convective overturn in the outer shells of the proto-neutron star (prompt convection). Prompt convection subsides quickly after the unstable region is mixed and neutrino losses subside.

\begin{figure}
    \centering
    \includegraphics[width=\linewidth]{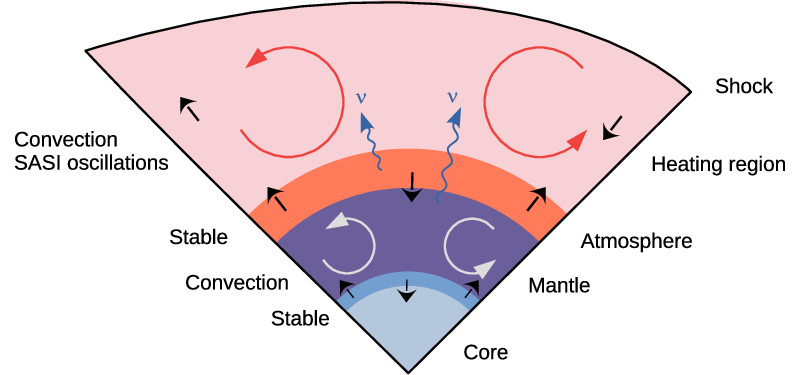}
    \caption{Sketch of the structure of the supernova core highlighting the instabilities and oscillatory motions that occur in different regions. Convective motions are denoted by circular arrows and oscillatory motions are denoted by short black arrows.  
    At the center of the proto-neutron star, there is a convectively stable low-entropy core (blue), surrounded by the convective mantle (indigo). Further out, there is a convectively stable atmosphere (brick red). A fraction of neutrinos from the atmosphere and the outer layers of the mantle deposit their energy in the heating region (light red). The heating region may be unstable to convection due to neutrino energy deposition, or to SASI (standing accretion shock instability) shock oscillations. The buoyancy frequency is particularly high in the proto-neutron star atmosphere and the steep entropy gradient in the outer region of the core (blue annulus).}
    \label{fig:sketch}
\end{figure}

\subsection{Post-bounce structure of the supernova core}
The structure of the supernova core during
the subsequent quasi-stationary accretion phase 
is sketched in Figure~\ref{fig:sketch}, with a particular emphasis on hydrodynamic instabilities that occur in various layers.
Starting from the interior, there is the high-density, low-entropy core of the proto-neutron star that was not subject to shock heating and is stably stratified. Further out, there is the proto-neutron star mantle with higher entropy, typically $5\texttt{-}6\,k_\mathrm{B}/\mathrm{nucleon}$. Due to energy and lepton number losses at the proto-neutron star surface, the mantle becomes and remains unstable to convection a few tens of milliseconds after bounce; the interplay  of entropy and lepton number gradients that drive convection and regulate its quasi-stationary state is quite complex \citep[e.g.,][]{bruenn_96,glas_19,jakobus_25}. Further out at lower densities, there is the proto-neutron star atmosphere,
where neutrino emission and absorption maintain a roughly isothermal and convectively stable stratification with a steep density ``cliff''.

Outside the proto-neutron star, a region of net neutrino energy deposition (gain region)  develops between the proto-neutron star atmosphere and the shock, typically about
$80\texttt{-}100\,\mathrm{ms}$ after bounce. Net heating in the post-shock accretion flow makes the gain region unstable to buoyancy-driven convection \citep{herant_94,burrows_95,janka_95}. However, the gain region may remain convectively stable if advection through the gain region is fast compared to the characteristic growth time scale for convective instability \citep{foglizzo_06}.

In addition to convection, the standing accretion-shock instability (SASI) may develop in the supernova core \citep{blondin_03,blondin_07}. The SASI is a large-scale oscillatory instability mediated by a feedback cycle of acoustic wave and vorticity waves between the shock and proto-neutron star surface \citep{blondin_06,foglizzo_07}. The SASI does not rely on neutrino heating and can also grow when fast advection stabilizes the gain region against convection \citep{mueller_12a,hanke_13}. Empirically, there seem to be two distinct regimes of slow and fast advection
in which neutrino-driven convection and
the SASI operate exclusively \citep{fernandez_14a,fernandez_15}.

Although convection and the SASI are the most prominent and dynamically important fluid instabilities in supernova of non-rotating or slowly rotating progenitors, further instabilities can occur. Already without rotation, dynamo amplification of magnetic field occurs and may provide some additional support for neutrino-driven explosions \citep{mueller_20b}. For rapid rotation, magnetic field amplification by the magnetorotational instability \citep{akiyama_03,obergaulinger_09} may occur in various regions of the supernova core, and inside the proto-neutron star other dynamo processes also come into play, especially on long time scales \citep{raynaud_20,barrere_22}.

Despite these multi-dimensional hydrodynamic instabilities, the accretion flow is quasi-stationary over longer time scales during the
pre-explosion phase.
Ultimately, the accretion flow through the shock
onto the proto-neutron star must be turned into an explosion in most massive
stars. There are several mechanisms for injecting energy into the post-shock region, driving the shock outward, and expelling matter in its wake, which we shall discuss in the remainder of this section. Among these,
the \emph{neutrino-driven mechanism} (Section~\ref{sec:sh_revival}--\ref{sec:phenomenology}) has been studied most extensively. The neutrino-driven mechanism poses no specific requirements in terms of special progenitor properties, and is therefore thought to be responsible for the vast majority of core-collapse supernova explosions. However, the neutrino-driven mechanism is likely not able to explain explosions far above $10^{51}\,\mathrm{erg}$, which are observed in ``hypernovae'' that constitute about $1\%$ of the core-collapse supernova
population in the local universe \citep{woosley_06b,smith_11}.
These are most likely explained by \emph{magnetorotational explosions},
which involve energy extraction from a rapidly rotating neutron star
or a black-hole accretion disk by magnetic fields (Section~\ref{sec:mrot}).
As the requisite rapid progenitor rotation for magnetorotational explosions is expected only in rare evolutionary channels
\citep[e.g.,][]{woosley_06b,yoon_06,podsiadlowski_10,aguilera_18}, magnetorotational explosions are not a viable option for explaining the bulk of the core-collapse supernova population.
Moreover, the initial angular momentum profiles, field strengths, and magnetic field geometries still constitute a major uncertainty in this scenario \citep{mueller_25b}.
A third scenario, the \emph{phase-transition mechanism} involves a second collapse and rebound of the proto-neutron
star to launch a second shock that explodes the star. This scenario remains more speculative and could only explain a fraction of all core-collapse supernovae (see Section~\ref{sec:pt} for details).

\subsection{Shock Revival -- Neutrino-Driven Mechanism}
\label{sec:sh_revival}
The neutrino-driven mechanism for shock revival has a long
history \citep{colgate_66}. The modern idea of ``delayed'' neutrino-driven explosions after an extended accretion phase was first enunciated by \citet{bethe_85}, and extensive theory has been developed to underpin it
since then. In the quasi-steady accretion flow, the shock position adjusts itself according to the thermal state of the gain region (which is influenced by neutrino heating), and the pre-shock conditions, which are set by the mass accretion rate. Once heating in the gain region becomes sufficiently strong, this balance can no longer be maintained, and runaway shock expansion ensues. More precisely, the time scale for unbinding the material in the gain region (heating time scale) becomes shorter than the advection time scale across the gain region to trigger \emph{neutrino-driven shock revival} \citep{janka_01,fernandez_12}. The multi-dimensional fluid flow in the gain region aids shock revival by pushing the shock further out via Reynolds stresses and energy redistribution \citep{murphy_12,mueller_15a,mabanta_18}, thus increasing the advection time and contributing to more favorable neutrino heating conditions.
The violence of the non-spherical mass motions in the heating region, quantified by the turbulent Mach number \citep{mueller_15a}, determines to what extent hydrodynamical instabilities facilitate shock revival.

The progenitor structure plays a key role in determining the explosion conditions, as it sets the mass accretion rate $\dot M$ onto the shock, and hence the pre-shock density and ram pressure. After the early post-bounce phase, $\dot M$ is related to the pre-collapse density profile $\rho(m)$, expressed as a function of mass coordinate $m$,
\begin{equation}
\dot{M}
\approx \frac{2m}{t_\mathrm{FF}}\frac{\rho(m)}{\bar\rho(m)}
\approx
\frac{8\rho}{3}\sqrt{3 Gm r^3},
\end{equation}
where $r$ is the radius of the mass shell in the progenitor, $t_\mathrm{FF}$ is the free-fall time scale for the shell and $\bar \rho(m)$ is the average density inside the shell \citep{mueller_16b}. A higher mass accretion rate implies a higher ram pressure and pre-shock density, but on the other hand it also leads to a higher accretion luminosity in neutrinos and hence stronger heating. Hence the competition between neutrino heating and the accretion ``tamp'' onto the shock is delicate and needs to be studied by numerical simulations. 
Details in the neutrino transport and neutrino microphysics \citep[e.g.,][]{lentz_12b,melson_15b}, the nuclear equation of state \citep[e.g.][]{janka_12b,suwa_13,yasin_20,ghosh_22,powell_25,rusakov_26}, and general relativistic effects \citep{mueller_12a,dunham_24} influence the proto-neutron star contraction, the emerging neutrino fluxes and spectra,
the development of hydrodynamic instabilities, and hence the explosion conditions.
Collective neutrino flavor conversion is an especially thorny issue, as this involves instabilities and equilibration processes on scales that cannot be resolved in global simulations \citep{tamborra_21}. There have been first efforts to incorporate them into global simulations using well-conceived effective models \citep{ehring_23,wang_25}, but there is still a long way towards a consistent treatment of quantum kinetics in supernova models \citep{mezzacappa_20}. Nonetheless, some general trends on explodability and explosion energetics are emerging from modern multi-dimensional simulations (see Section~\ref{sec:phenomenology}).

In recent years, it has also been recognized that pre-collapse asphericities in supernova progenitors can play an important role in the development of neutrino-driven explosions \citep{couch_13,mueller_15a,couch_15,mueller_17}. Convective motions in the burning shells around the iron core of moderate Mach numbers
of order $\mathrm{Ma}\sim0.1$ will translate into sizable pre-shock density perturbations during collapse \citep{mueller_15a,abdikamalov_19}, facilitating asymmetric shock expansion and injecting turbulent energy into the gain region \citep{mueller_16c,abdikamalov_18}. While some supernova simulations with multi-dimensional progenitor models have been performed \citep{couch_15,mueller_17,bollig_21}, stronger integration of hydrodynamical simulations of the pre-collapse stages and the subsequent collapse and explosion is still required.

\subsection{Brief Phenomenology of 3D Supernova Explosion Simulations}
\label{sec:phenomenology}
As discussed in the Introduction, it is beyond the scope of this review to exhaustively catalog the current state of supernova explosion simulations, and I refer to dedicated reviews for this purpose \citep{mueller_20,mezzacappa_20,burrows_21,janka_25}. I rather focus on a high-level summary of the developments pertinent to supernova GW astronomy. Three-dimensional simulations of successful explosions based on multi-group neutrino transport have now been available for about a decade \citep{lentz_15,melson_15a,melson_15b,takiwaki_14}.
Since then, models have matured in many respects. For GW astronomy, the principal improvement may have been one of scale -- a larger number of longer simulations, sometimes beyond the first second of the explosion \citep{mueller_15b,mueller_17,burrows_20,bollig_21,mezzacappa_23,janka_24}. Hence, a larger set of longer and less incomplete set of waveform predictions is now available for testing classification and parameter inference techniques using mock data.

The 3D simulations from the last decade have yielded key insight into the neutrino-driven mechanism. They have demonstrated that the neutrino-driven mechanism can roughly achieve typical supernova explosion energies with a relatively long power-up phase that can last several seconds \citep{mueller_17,burrows_20,bollig_21}. They naturally produce a reasonable range of neutron star kicks and neutron star birth spin rates \citep{mueller_19a,burrows_24a,janka_24},
even without assuming any progenitor rotation. There is a tendency towards higher explosion energies and neutron star kicks for explosions originating from progenitors with more massive cores \citep{mueller_19a,burrows_24a}. 
Modern 3D simulations have also proved useful in explaining unexpected findings about the compact object mass distribution. For example, the detection of a $\sim 2.8\,\msun$ black hole in the GW merger event GW190814, in what was previously thought to be a ``mass gap'' between neutron stars and black holes \citep{belczysnki_12} is naturally accounted for by fallback supernovae with black-hole formation in a successful explosion. 

However, to date no simulations can as yet fully take into account all the aforementioned factors (Section~\ref{sec:sh_revival}) that affect the chance of shock revival and the explosion and remnant properties in case of a successful explosion. Modern 3D simulations show -- by and large -- a consensus that neutrino-driven explosions are achievable, can reach plausible explosion and remnant properties and hint at correlations of these properties with the progenitor structure. But ab-initio modeling has not yet reached a consensus on the progenitor mass range for successful explosions (with neutron star or possibly sometimes black hole formation), although explosion models
for some of the least massive progenitor models appear more robust \citep{melson_15a,mueller_16b,sandoval_21,zha_22,wang_24,mueller_25}. This does not imply complete uncertainty either, especially when simulations are combined with some form of (implicit) calibration or validation against observational constraints.

\subsection{Shock revival -- Magnetorotational mechanism}
\label{sec:mrot}
For rapid rotation, the rotational energy of the proto-neutron star, or perhaps an accretion disk around a black hole at a later stage provides another energy reservoir that can be tapped by magnetic fields to power a \emph{magnetorotational} explosion.
The maximum (Keplerian) spin-rate for rigidly rotating cold neutron stars corresponds to a rotational energy of about $10^{52}\,\mathrm{erg}$. This roughly matches the highest inferred explosion energies of broad-lined Ic supernovae or ``hypernovae'' \citep{woosley_06b}, which has been taken to suggest rapidly rotating \emph{millisecond magnetars} as the engine behind such explosions \citep{uzov_92,duncan_92,mazzali_14}. Black hole accretion disks in the \emph{collapsar scenario} \citep{macfadyen_99} provide an energy reservoir of similar, perhaps even larger magnitude. Although neutrino-driven explosions are ultimately fed by the large reservoir of thermal energy of $\gtrsim 10^{53}\,\mathrm{erg}$ in the proto-neutron star, the neutrino-driven mechanism can only transfer $\sim 10^{51}\,\mathrm{erg}$ into kinetic energy of the ejecta and is therefore unlikely to explain hypernovae.

Magnetohydrodynamic effects also more naturally explain collimated bipolar outflows as the main driver of the explosion or an accompanying component. This fits the observational evidence on hypernovae, where bipolar outflows are indicated by polarization measurements
and there is often an associated long gamma-ray burst \citep{woosley_06b} (though the relativistic gamma-ray burst jets are a distinct phenomenon from the outflows that contain the bulk of the ejecta mass and energy).

Simulations of magnetorotational explosions have matured considerably in recent years. However, while three-dimensional magnetohydrodynamic simulations are now routinely conducted \citep[e.g.,][]{winteler_12,moesta_14b,kuroda_20,obergaulinger_21,powell_23}, some even with multi-group neutrino transport, these are subject to more substantial uncertainties than neutrino-driven explosion models. Pathways towards the evolution of the requisite rapidly rotating progenitors are uncertain due to our incomplete understanding of rotation and magnetic fields in stellar evolution. The pre-collapse magnetic fields are a major unknown,
and are only now being investigated with the help of 3D simulations of magnetoconvection \citep{varma_23b}. Magnetic field amplification after collapse is complex and involves small-scale processes \citep{obergaulinger_09,guilet_15,guilet_15b,masada_12,moesta_15} that cannot be resolved in global multi-physics simulations over longer time scales. Because of such complications, the systematics of explosion and remnant properties are not yet understood to the same degree as for neutrino-driven explosions.

\subsection{Explosions triggered by a phase transition?}
\label{sec:pt}
Finally, it has been suggested that some core-collapse supernova explosions may be triggered by a phase transition to deconfined quark matter in the proto-neutron star. A first-order phase transition from nucleonic to quark matter could trigger another collapse of the proto-neutron star to a more compact configuration, followed by a second bounce and the launching of a powerful shock wave \citep{fischer_10,fischer_18}. This mechanism is, however, sensitive to details of the transition to quark matter, and nuclear physics constraints imply that this mechanism cannot operate generically in explosions that make neutron stars of typical masses around $1.4\msun$. Even with a first-order phase transition, the mass range and achievable energies for explosion triggered by a phase transition remain subject to debate
\citep{zha_21,jakobus_22}. However, a phase transition could produce interesting GW signals, regardless of whether it is a first-order phase transition or a smooth crossover (see Section~\ref{sec:gw_pt}).

\section{The Supernova Gravitational Wave Signal}
\label{sec:gw}
\subsection{Basic Considerations}
Core-collapse supernovae emit GWs because of the non-spherical mass motions caused by  (magneto-)hydrodynamic instabilities and bulk rotation (if present). In addition, asymmetric neutrino emission also leads to GW emission \citep{epstein_78}.
Before considering the phenomenology and detailed theory of the predicted GW signals, it is useful to establish some rough dependencies on physical parameters.

Since core-collapse supernovae are only mildly relativistic systems in most circumstances, the GW strain tensor $h_{ij}^\mathrm{TT}$ from aspherical mass motions is (to first approximation) given by the Einstein quadrupole formula \citep{einstein} in the transverse-traceless (TT) gauge,
\begin{equation}
\label{eq:quadrupole}
h_{ij}^\mathrm{TT}
=\frac{2 G}{c^4 D} \mathrm{STF}(\ddot{Q}_{ij}),
\end{equation}
where $D$ is the distance from the source, $Q_{ij}$ is the mass quadrupole moment, and STF denotes the projection onto transverse-traceless components. The second time derivative of the quadrupole component can be re-expressed using the conservation laws for energy and momentum and integration by parts, yielding the stress formula \citep{nakamura_89,blanchet_90}
in terms of the density and velocity fields $\rho$ and
$v_i$ and the gravitational potential $\Phi$,
\begin{equation}
\label{eq:stf}
h_{ij}^\mathrm{TT}
=\frac{2 G}{c^4 D} \mathrm{STF}\left[\int \rho (v_i v_j-x_i\partial_j\Phi)\,\ud V\right].
\end{equation}

Equation~(\ref{eq:stf}) offers insights into the factors that determine the GW amplitude. More mass involved in asymmetric mass motions and larger non-spherical velocities will increase the first term in the integral. The second term is largely sensitive to strong asphericities in the density distribution and again to the mass involved. In the case of oscillating fluid motions, the two terms in the integral are somewhat related to the kinetic energy and the potential energy in the oscillations, although of course only the quadrupolar component of such fluid motions sources GWs. In the case of oscillating fluid motions, we expect the two terms in the integral to be of a similar magnitude.

One can express this idea more formally to estimate typical GW amplitudes by dimensional analysis.
In terms of the mass  $M$ and kinetic $E_\mathrm{kin}$
involved in aspherical motions, the Schwarzschild radius $R_\mathrm{s}$ corresponding to $M$, the characteristic length scale of fluid motions $R$, the angular frequency $\omega$, and aspherical velocities $v$ one has
\begin{equation}
h_{ij}^\mathrm{TT}\sim \frac{2\alpha_\mathrm{Q}GE_\mathrm{kin}}{d c^4}
\sim \frac{\alpha_\mathrm{Q}}{d}\frac{2G M}{c^2}\left(\frac{v}{c}\right)^2
\sim
\frac{\alpha_\mathrm{Q}}{d}\frac{2G M R^2 \omega^2}{c^2}
\sim 
\frac{\alpha_\mathrm{Q} R_\mathrm{s}}{d}
\left(\frac{v}{c}\right)^2.
\end{equation}
He the dimensionless factor $\alpha_\mathrm{Q}$ 
accounts for the fact that only part of the energy in oscillating fluid motions is actually contained in quadrupole modes that contribute to GW emission.

For anisotropic neutrino emission, the slow-motion approximation inherent in the classical quadrupole formula is inappropriate, but 
the amplitude can still be derived in the weak-field approximation using retarded Green's functions
\citep{epstein_78}. The gravitational wave strain can be expressed quite conveniently in terms of the emerging neutrino energy flux per solid angle , $\ud L_\nu (\mathbf{n}',t')/\ud \Omega'$, as a function of direction $\mathbf{n'}$ and time $t'$
\citep{mueller_97},
\begin{equation}
\label{eq:gw_nu}
h_{ij}^\mathrm{TT}  = \frac{ 4G }{ c^4D} \int_{-\infty}^{t-D/c}  \int_{4\pi}
\frac{ (n_i n_j)^\mathrm{TT} }{ 1-\mathbf{n}\cdot\mathbf{n}' } \cdot 
\frac{\ud L_{\nu}(\mathbf{\mathbf{n'}},t') }{\ud\Omega'} \,\ud\Omega' \,\ud t',
\end{equation}
where $\mathbf{n}$ is the direction of the observer.
Note that the neutrino energy flux is the total flux for neutrinos and antineutrinos of all flavors here.
The time integral in 
Equation~(\ref{eq:gw_nu}) implies a \emph{memory effect}: Anisotropic neutrino emission produces a strain component that does not return to zero after the event.
In terms of the neutrino luminosity $L$ and the total radiated energy $E_\nu$ in all directions, we thus have
\begin{equation}
    \frac{\ud h_{ij}^\mathrm{TT}}{\ud t}
    \sim \frac{\alpha_\nu G L}{c^4 D},
    \quad
    h_{ij}^\mathrm{TT}
    \sim \frac{\bar\alpha_\nu G E_\nu}{c^4 D},
\end{equation}
with appropriate dimensionless factors $\alpha_\nu$
and $\bar\alpha_\nu$ to characterize the instantaneous and time-averaged neutrino emission anisotropy.
Here $\alpha_\nu$ only measures those large-scale emission anisotropies that  contribute to GW emission in the far field.
Note that $\bar{\alpha}_\nu$ is obviously not a simple (weighted)
time average of $\alpha_\nu$ since
 the directional dependence of
the neutrino emission anisotropy varies over time.
It characterizes both the overall degree of emission anisotropy as well as the temporal stability of the directional dependence.

In all of these expressions, the superscript $\mathrm{TT}$ denotes the symmetric transverse-traceless component, which is
obtained using the transverse projection tensor
$P_{ik} =\delta_{ik}-n_i n_k$. For any rank-2 tensor $X_{ij}$,
\begin{equation}
    X_{ij}^\mathrm{TT} =P_{ik} P_{jl} X^{kl}-\frac{1}{2}
    P_{ij}P_{kl} X_{lk},
\end{equation}
in Euclidean space in Cartesian coordinates as appropriate in the far-field regime.

As core-collapse supernovae are only mildly relativistic systems with comparably weak wavelike perturbations in the space-time, direct extraction of GWs is generally impractical in fully general relativistic simulations. The Newtonian quadrupole formula or its reformulations, the time-integrated quadrupole formula or stress formula \citep{finn_90,nakamura_89,blanchet_90} are therefore used for GW extraction, with appropriate correction terms in relativistic simulations.

\begin{figure}
    \centering
    \includegraphics[width=\linewidth]{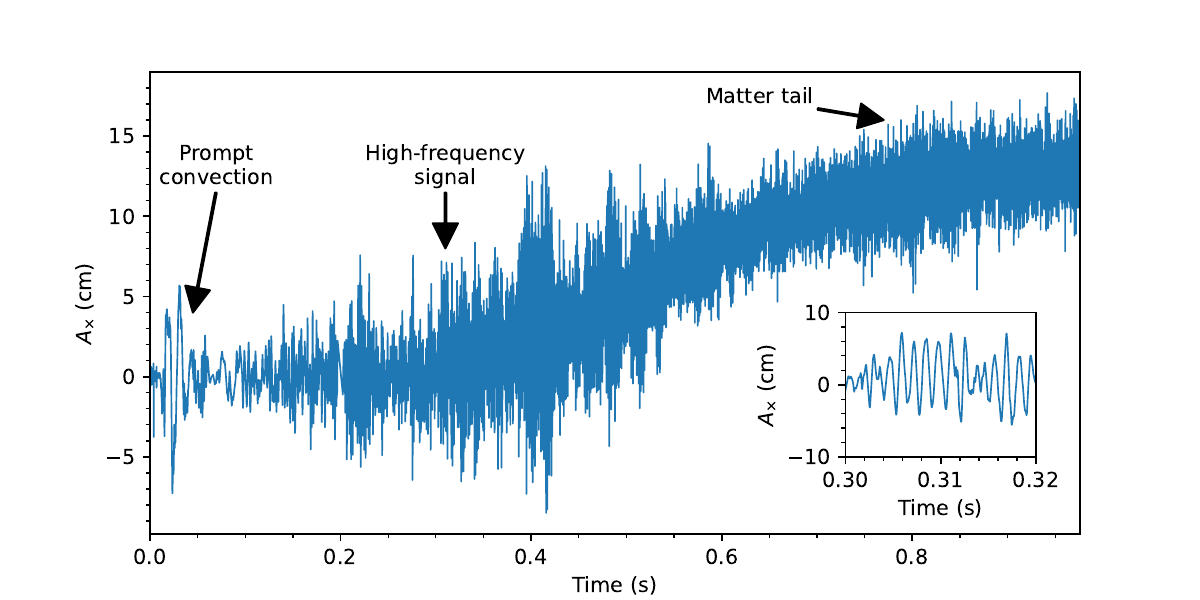}
    \caption{Illustration of the typical structure of the core-collapse supernova GW signal from a relativistic 3D simulations of a non-rotating $20\msun$ star \citep{yoon_17}. The plot shows the distance-normalized amplitude $A_\times$ for one observer direction.
    The first convection is a signal from the shock and proto-neutron star oscillations triggered by prompt convection. This is followed by a more quiet period, until the high-frequency signal develops. The inset shows that this signal has a clear quasi-periodic structure despite the stochastic amplitude modulations. After the explosion, asymmetric shock expansion produces a slowly growing offset in the GW signal. Note that the neutrino memory signal, which typically dominates the matter memory signal, is not shown here.}
    \label{fig:gw_template}
\end{figure}

\subsection{Phases of Gravitational Wave Emission in Core-Collapse Supernovae}
\begin{figure}
    \centering
    \includegraphics[width=\linewidth]{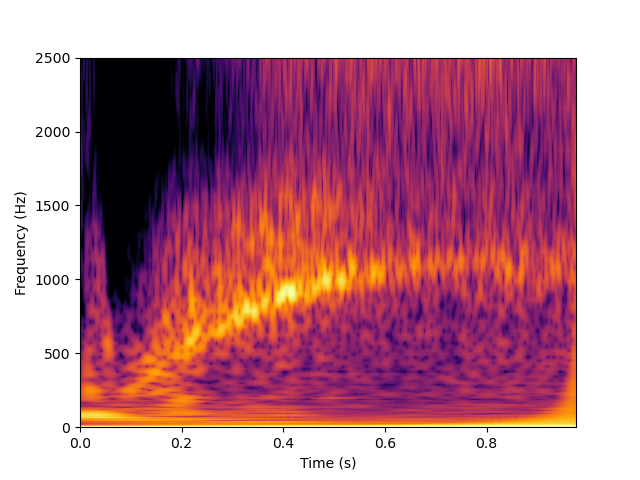}
    \caption{Typical spectrogram of a predicted core-collapse supernova GW signal, from the same $20\msun$ model as shown in Figure~\ref{fig:gw_template}. The most prominent feature in the spectrogram is the high-frequency ramp-up signal, which rises from a few hundred Hz to above 1\,kHz in this simulation. There is some haze of $p$-modes above this dominant emission band, in particular when the GW amplitudes are highest. The signal from prompt convection contributes some power at frequencies of $\sim 100\,\mathrm{Hz}$ at early times.}
    \label{fig:spec}
\end{figure}

It is useful to provide a rough overview of key ``standard'' features of the supernova GW signal 
in the time domain (Figure~\ref{fig:gw_template})
and time-frequency (Figure~\ref{fig:spec}) domain from the collapse to the explosion phase as seen in modern 3D simulations of neutrino-driven explosions.
I will then review the underlying theory and also consider less universal features.

For sufficiently rapid rotation, the first signal component is the rotational bounce signal. The bounce signal comes from 
the contraction, rebound, and ringdown of the rotationally deformed inner core, and had long been a major focus of core-collapse supernova GW theory. However, over the last decades, stellar evolution predictions have settled towards relatively modest spin rates of the progenitor cores \citep{heger_05,ma_19}, partly prompted by observational constraints on angular momentum transport from asteroseismic measurements for evolved low-mass stars \citep{mosser_12}. In typical core-collapse supernovae, this component is therefore likely absent or not prominently visible next to the subsequent GW emission.

For the more generic case of non-rotating or slowly-rotating progenitors, GW emission starts shortly after bounce with a signal triggered by prompt convection. This signal can persist at low frequencies of $\lesssim 100\,\mathrm{Hz}$ and declining amplitudes for tens of milliseconds after convection subsides. This longer emission comes from shock oscillations kicked off by prompt convection and the associated propagation of acoustic waves between the shock and the proto-neutron star \citep{mueller_13}.

This is typically followed by a quieter period of tens of milliseconds, until the main signal components develop in earnest. The most prominent and robust of these is a high-frequency emission band (``ramp-up signal'') that rises from about $200\,\mathrm{Hz}$ to  $\gtrsim 1\,\mathrm{kHz}$ during
the first second after bounce. With an appropriate time-frequency analysis, this signal usually emerges as quite sharply defined. Another prominent component occurs in simulations that show SASI activity \citep{kuroda_16b,andresen_17,hayama_18,mezzacappa_20b,powell_21}. This SASI-associated component lives at lower frequencies, typically in the range $\lesssim 200\,\mathrm{Hz}$. It often has a more complex frequency trajectory and is more intermittent than the
dominant high-frequency emission band. Finally, asymmetric
shock expansion \citep{murphy_09,mueller_13} and anisotropic neutrino emission \citep{epstein_78,marek_08,yakunin_16} produce a tail signal after the onset of explosion, which only contributes at very low frequencies. 
Asymmetric neutrino emission usually contributes far more strongly to the tail than the matter signal. It generally produces significant emission in the low-frequency range
in the regions of tens of Hz and below even before the onset of explosion. 

Emission in the high-frequency band typically peaks during
the first few hundred milliseconds after shock revival and then continues at a lower level. The high-frequency band tends
to be stronger in exploding models than in non-exploding ones,
and also tends to be stronger in more massive progenitors
both in 2D and 3D \citep{mueller_13,radice_19}.
The reason for this correlation will become clear from a closer analysis of the physics underlying this signal component (Section~\ref{sec:hf_signal}).
The SASI signal stops once the shock is revived \citep{andresen_17}. It tends to last longer and appear more frequently and cleanly in spectrograms of non-exploding models.

A summary of key features of the supernova GW signal is provided
in Table~\ref{tab:sig_comp}. The table shows typical amplitudes, frequency ranges and durations for the main signal components, and the conditions under
which these component appear.

\begin{table}[]
    \centering
    \begin{tabular}{ccccc}
    \hline\hline
        Signal component & $h D$& Frequency & Duration  & Conditions\\
        \hline
         Rotational bounce & $\lesssim 100\,\mathrm{cm}$ & $\mathord{\sim}700\,\mathrm{Hz}$ & a few ms & Rapid rotation\\
         Prompt convection & $\mathcal{O}(1 \,\mathrm{cm})$ & $\mathord{\sim}
         100\,\mathrm{Hz}$ & tens of ms & ---\\
         Ramp-up signal & $\mathcal{O}(1\texttt{-}10)\,\mathrm{cm}$ & $
         \mathord{\sim}0.5\texttt{-}1.5\,\mathrm{kHz}$ & $\mathcal{O}(0.1\texttt{-}1\,\mathrm{s})$ & ---\\
         SASI &  $\lesssim  1\,\mathrm{cm}$ & $100\texttt{-}200\,\mathrm{Hz}$ &
        $\mathcal{O}(0.1\texttt{-}1\,\mathrm{s})$ & Before explosion$^\dagger$\\
         Neutrino memory & $\mathcal{O}(10^{2\texttt{-}3}\,\mathrm{cm})$ &
         mostly $\lesssim 10\,\mathrm{Hz}$& seconds & ---\\
         Triaxial modes & $\mathcal{O}(100\,\mathrm{cm})$ &
         a few $100\,\mathrm{Hz}$ & up to  $\gtrsim 0.1\,\mathrm{s}$ &
         Rapid rotation$^\dagger$\\
         \hline\hline
    \end{tabular}
    \caption{Summary of key signal components of supernova GW
    signals showing typical values for the distance-normalized strain $hD$, the frequency, and the signal duration under the conditions specified
    in the last column. If the condition is marked with a dagger ($\dagger$), the associated signal may not always appear, or may appear intermittently. Note that there is substantial variability among predicted supernova GW signals. Values outside the indicated ranges may be encountered. For other, less commonly studied signal components refer to the text.}
    \label{tab:sig_comp}
\end{table}

\subsection{The Bounce Signal}
Starting with the signal from rotational core bounce, let us now consider more closely the principal characteristics of the aforementioned signal components and the physical dependencies that govern them. Upon the inclusion of deleptonization during the collapse stage, the bounce signal was found to be of highly uniform shape except for extremely high rotation rates \citep{dimmelmeier_07_a,dimmelmeier_08}.
The peak frequency of the time-integrated spectrum 
is around $700\,\mathrm{Hz}$ and
only moderately depends on the progenitor mass and rotation rate and the nuclear EoS\citep{dimmelmeier_07_a,dimmelmeier_08}.
The uniformity across progenitors comes from the fact that deleptonization brings the final mass of the homologous inner core at bounce to very similar values with little sensitivity to the initial iron core mass. The bounce signal is principally generated by the fundamental oscillation mode ($f$-mode) of the proto-neutron star, which implies a scaling
of the mode frequency $f$ with central density $\rho_c$ as
$f\sim (2\pi)^{-1}\sqrt{G\rho_c}$ \citep{fuller_15b}.

Since the quadrupolar deformation parameter $\alpha_Q$ depends on the ratio of the angular velocity $\Omega$ to 
the dynamical frequency $\Omega_\mathrm{dyn}$ roughly as
$\alpha_Q\sim(\Omega/\Omega_\mathrm{dyn})^2=\Omega^2R^3/(GM)$
\citep{fuller_15b}, the amplitude of the bounce signal 
is tightly related to the initial ratio of rotational energy $T$ and binding $|W|$ of the iron core up
to initial values of $T/|W|\sim 0.07$ \citep{abdikamalov_14}.
This tight dependence and the templatability of the bounce
signal provide the opportunity to quantitatively constrain
progenitor rotation in the event of Galactic supernova.
However, a rotational bounce signal will only provide a minimum bound
on  $T/|W|\sim 0.07$ since the measurable strain $h$ depends on the angle $\theta$ between the rotation axis of the core and the observer direction,
$h\propto \sin \theta$. The ambiguity from the orientation of the
rotation axis could potentially be broken if there is evidence of a
bipolar magnetorotational explosion and electromagnetic observations can constrain the symmetry axis of the ejecta.

\subsection{The Prompt Convection Signal}
After bounce, aspherical mass motions that source gravitational waves also arise in the absence of rotation. The first phase of substantial GW emission in the case of slow or negligible rotation is triggered by prompt convection (Section~\ref{sec:bounce}). The convection itself quickly mixes the unstable region and then subsides, and the actual GW emission from this early phase predominantly comes from oscillations of the shock and post-shock matter that are excited by the convective plumes \citep{mueller_13}. This results in a quasi-periodic signal with a frequency of $\sim100\,\mathrm{Hz}$. The amplitude of this early periodic signal varies considerably across simulations. The width of the unstable convective region, the detailed entropy gradient left by the shock, and the violence and scale of the turnover motions determine the amplitude of the quadrupolar oscillations excited by prompt convection. These factors are influenced by details of the neutrino transport and interaction rates and by general relativistic effects \citep{mueller_10}, and possibly by physical and numerical seed perturbations. By and large, the early prompt convection signal also tends to be smaller in 3D simulations than in 2D.

\subsection{The High-Frequency Signal}
The high-frequency ``ramp-up'' signal tends to be the most robust feature of GW emission in core-collapse supernovae. The physics behind this signal first emerged in 2D simulations of substantial duration. 

\subsubsection{Frequency structure}
Murphy et al.\ \citep{murphy_09} identified the critical role of buoyancy forces in the convectively stable proto-neutron star surface in shaping the signal. Using a wavelet analysis of the signal, M\"uller et al.\ \citep{mueller_13} demonstrated the presence of a clearly identifiable narrow emission band rather than a completely stochastic signal. The well-defined frequency trajectory points to a specific proto-neutron star oscillation mode that is excited by the aspherical fluid motions next to the stable proto-neutron star surface region. Using a local approximation for the frequency of buoyancy-dominated modes ($g$-modes) leads
to the frequency relation \citep{mueller_13}
\begin{equation}
  f_\mathrm{g}
  \approx
\frac{1}{2\pi}
  \frac{GM}{R^2}
\sqrt{1.1 \frac{m_\mathrm{n}}{\langle E_{\bar{\nu}_e}\rangle}}
\left(1-\frac{GM}{Rc^2}\right)^{2},
\label{eq:fpeak}
\end{equation}
where $M$ and $R$ are the proto-neutron star mass and radius (measured at a density of $10^{11}\,\mathrm{g}\,\mathrm{cm}^{-3}$), $m_\mathrm{n}$ is the neutron mass and the mean energy $\langle E_{\bar{\nu}_\mathrm{e}}\rangle$ of electron antineutrinos serves as a proxy for the proto-neutron star surface temperature. In other words, the frequency is determined by the proto-neutron star surface gravity and surface temperature, and there is also a relativistic correction factor that accounts for time dilation and other relativistic effects in the linearized equations of fluid dynamics.
The important implication is that the frequency trajectory, if measurable, holds direct information on the time-dependent proto-neutron star properties, i.e., its accretion history and its contraction.

The high-frequency mode is the most prominent proto-neutron star oscillation mode visible in the spectrogram, but not the only one. Linear perturbation theory \citep{sotani_16,torres_18,torres_19,morozova_18} provides a useful tool for studying the structure of these modes more rigorously. This leads to an eigenvalue problem for the radial and transverse displacements $\eta_r$ and $\eta_\perp$. In the Cowling approximation, i.e., neglecting space-time perturbations, the eigenvalue problem for adiabatic oscillations 
with frequency $\sigma$ in a conformally flat space-time reads \citep{torres_18},
\begin{eqnarray}
\partial_r \eta_r 
+ \left[\frac{2}{r} + \frac{1}{\Gamma}\frac{\partial_r P}{P} + 6\frac{\partial_r \psi}{\psi}\right]\eta_r
+ \frac{\psi^4}{\alpha^2 c_s^2}\left(\sigma^2 - \mathcal{L}^2\right)\eta_\perp &=& 0, 
\\
\partial_r \eta_\perp 
- \left(1 - \frac{\mathcal{N}^2}{\sigma^2}\right)\eta_r
+ \left[\partial_r \ln (\rho h \alpha^{-2} \psi^4) - G\left(1 + \frac{1}{c_\mathrm{s}^2}\right)\right]\eta_\perp &=& 0,
\end{eqnarray}
where $\Gamma$ is the adiabatic index, $P$ is the pressure, $\rho$ is the density, $h$ is the relativistic enthalpy, $c_\mathrm{s}$ is the sound speed and $\alpha$ and $\psi$ are the lapse function and conformal factor. Finally,  $\mathcal{N}$ and $\mathcal{L}$ are the relativistic Brunt-V\"ais\"al\"a and Lamb frequency. 

The eigenfunctions permit a classification into $p$- (pressure) and $g$- (gravity) modes and the fundamental $f$-mode  according to the dominant restoring forces. 
Note that different classification schemes exist \citep{rodriguez_23}, which may lead to confusion in comparing different studies. Additional modes can occur in the presence of rotation and magnetic fields, but are typically not addressed by current codes for linear modes analysis for proto-neutron stars. Further subtleties include the boundary conditions at the shock \citep{sotani_19,westernacher_20}, deviations from the assumption of a hydrostatic background structure due to accretion through the shock \citep{tseneklidou_25}, and non-adiabaticity.

Linear mode analysis has revealed the high-frequency mode to track the $f$-mode or a low-order $g$-mode \citep{torres_18,morozova_18}. The high-frequency signal often switches between two branches of the mode spectrum \citep{radice_19,murphy_25}, but it fundamentally remains a buoyancy-dominated mode mostly confined to the proto-neutron star surface region.

While perturbation theory is useful for precisely identifying the physical nature of the modes seen in the spectrogram, the inference of proto-neutron star and supernova core properties in the event of future GW observations will need to rely on simpler, calibrated relations for mode frequencies. Several semi-empirical relations for the mode frequencies have been proposed based simulations to express
the mode frequency directly in terms
of $M/R^2$ \citep{torres_19b} or $M/R^3$ \citep{sotani_21,sotani_24} to obviate the need for information about the neutron star surface temperature from neutrinos in Equation~(\ref{eq:fpeak}). Based on such ``universal'' relations, \emph{time-dependent} measurements of proto-neutron star parameters from noisy signals become possible at signal-to-noise ratios of $\sim 25$ \citep{powell_22}.

To what extent these frequency relations are truly universal requires scrutiny. Deviations have, for example, been reported in the case of very rapid rotation \citep{powell_20,powell_23,sykes_26} because angular momentum gradients modify the buoyancy frequency in the proto-neutron star surface region.

Through its dependence on the proto-neutron star radius, the trajectory of the high-frequency signal may provide clues about the nuclear equation of state \citep{eggenberger_21,wolfe_23,murphy_24}. However, in attempting to constrain the proto-neutron star structure and possibly nuclear physics, one also needs to bear in mind the sensitivity of the high-frequency signal to other modeling assumptions. A pseudo-Newtonian treatment of gravity incurs errors of $\sim 15\%$ compared to relativistic simulations. The approximation of monopole gravity also introduces a bias in frequency \citep{sotani_25}. The sensitivity to the treatment of neutrino transport is less well explored.

\subsection{Excitation Mechanisms and Amplitude 
of the High-Frequency Signal}
\label{sec:hf_signal}
The power of the high-frequency signal potentially contains information about the excitation processes that drive the oscillation mode underlying this signal component. Oscillations could  be driven by fluid motions \emph{outside} the proto-neutron star in the gain region, or by the convective motions \emph{inside}, or by a combination of both. The strength of the high-frequency signal could provide clues about the violence of the fluid motions in these unstable regions.

Pinpointing the dominant excitation mechanism -- and hence the physical information that can be distilled from a prospective measurement of the signal power -- is more subtle than understanding the frequency structure, however. Clues come from various forms of circumstantial evidence, and one should note that the relative importance of each driving mechanism may subtly depend on the physics assumed by different simulation codes.

\emph{Temporal correlations} between the power in the high-frequency signal and the driving instabilities constitute one means of determining  causation. Both parameterized 3D models \citep{mueller_e_12} as well as early 3D models with multi-group neutrino transport
\citep{andresen_17} suggested that both  aspherical motions in the gain region and proto-neutron star convection are important. These early 3D studies already showed that amplitudes typically peak during the early explosion phase when the turbulent kinetic energy and turbulent Mach number in the gain region is highest, and then persist at a lower level, which suggests a greater role for excitation of modes from outside during peak emission, and proto-neutron star convection as the primary driver of late GW emission.

Complementary evidence comes from the spatial structure of GW emission. In a \emph{regional analysis} \citep{andresen_17,mezzacappa_23,murphy_25}, one restricts the integration in the time-integrated quadrupole formula or stress formula to sub-domains, e.g., for some region~I, Equation~(\ref{eq:stf}) would reduce to
\begin{equation}
h_{ij,\mathrm{I}}^\mathrm{TT}
=\frac{2 G}{c^4 D} \mathrm{STF}\left[\int_{V_\mathrm{I}} \rho (v_i v_j-x_i\partial_j\Phi)\,\ud V\right],
\end{equation}
and the total amplitude is obtained by summing over different regions. This approach entails some pitfalls.\footnote{In the fully covariant view of space-time, GWs and their sources can of course not unambiguously localized anyway, but relatively weak fields and the asymptotically flat structure of space-time in core-collapse supernovae render this problem secondary.} Different from the amplitudes, the GW power cannot simply be obtained by summing up the contributions from different regions due to interference terms. Moreover, correctly treating surface terms for each region requires care \citep{zha_24b}. Finally, oscillation modes with eigenfunctions mostly localized in the proto-neutron star surface region will still extend into neighboring regions, so that interpreting the dominant ``source'' region as the driver of mode excitation is not trivial. 
Careful consideration of the strain contributions from different regions, combined with knowledge of the mode eigenfunctions and their emission efficiency, and of the spectrograms from different regions can provide significant insights \citep{murphy_25}.
 The latest results of Murphy et al.\ \citep{murphy_25} show that even though the GW-emitting modes involve significant mass motions throughout the proto-neutron star, there is a shift in the character of emitting modes and regions shift with time. Their findings indicate a stronger role for accretion onto the proton-neutron star as a driver of mode excitation early on and a larger role for driving by proto-neutron star convection at later times
 \citep{murphy_25}.

 Regional analyses can be refined to a space-frequency analysis relying on short-time Fourier transforms of the quadrupole component of the mass motions as a function of radius \citep{jakobus_23,zha_24c}. This allows a more direct identification of the emitting modes with those computed by linear analysis. When the mode is strongly localized, as in the case of the \emph{core} $g$-mode (see Section~\ref{sec:core_g}), this also helps identify the drivers of the mode.

Finally, quantitative arguments can be made about the relation between the violence of the forcing motions and the power in the high-frequency signal. This is both relevant for pinpointing the driving mechanism and the interpretation of the signal strength in the event of a detection.
Using analytic theory for the excitation of gravity waves by turbulent motions \citep{goldreich_90,lecoanet_13}, one can relate the mode energy $E_\mathrm{g}$ in steady state to the turbulent luminosity $L_\mathrm{turb}$, turnover time $\tau$, and Mach number $\mathrm{Ma}$ of the driving motions \citep{mueller_17},
\begin{equation}
    E_\mathrm{g}\sim \alpha\, \mathrm{Ma}\, \tau L_{\mathrm{turb}},
\end{equation}
where $\alpha$ is a factor that captures the overlap of the driving motions with the frequency and angular dependence (quadrupolar in the case of GW-emitting modes) of the excited mode. This factor is likely not too far below unity for large-scale driving motions. Relating the mode energy to the strain and energy of the emitted GWs leads to the estimate 
\begin{equation}
\label{eqegw}
    E_{\mathrm{GW}} \approx
4.2 \times 10^{43}\ \mathrm{erg}
\left(\frac{f}{1000\ \mathrm{Hz}}\right)^2
\left(\frac{\tau}{20\ \mathrm{ms}}\right)^2
\left(\frac{\mathrm{Ma}}{0.3}\right)^2
\left(\frac{E_{\mathrm{turb}}}{10^{50}\ \mathrm{erg}}\right)^2,
\end{equation}
for the total GW energy $E_\mathrm{GW}$ in terms of $\tau$, $\mathrm{Ma}$, and the time-integrated turbulent luminosity \citep{powell_19}. Among the various factors, the $E_\mathrm{turb}$ is mostly responsible for the progenitor dependence of the radiated GW energy. Systematic studies based on 3D explosion models suggest that the total GW emission roughly scales with the time-integrated turbulent luminosity in the \emph{gain region} with an effective power law of $h\propto E_\mathrm{turb}^{1.88}$ \citep{radice_19}. Strictly speaking, such a scaling with the time-integrated turbulent luminosity only suggests that the bulk of the GW emission comes from excitation of proto-neutron star oscillation by motions in the gain region (though this may be the primary question of interest from the observational point of view anyway). Furthermore, scaling relations based on analytic theory for mode excitation tend to give relatively similar estimates for excitation by motions in the gain region (with high Mach number, but lower energy) and by proto-neutron star convection (with low Mach number and higher energy). Thus, a synoptic view of the evidence related to the driving of oscillation modes remains necessary; no single analysis approach provides a definitive answer on its own.

Independent of the details of mode excitation, the growing number of 3D simulations has revealed a general trend towards stronger GW signals from more massive stars, and towards stronger signals from exploding ones than from non-exploding ones
\citep{radice_19}. This effectively confirms the trend already anticipated by 2D simulations \citep{mueller_13,yakunin_16}, though the predicted amplitudes in 3D are generally lower than in 2D.
The theory of mode excitation naturally explains this empirical dependence.
More massive progenitors tend to be characterized by higher accretion rates and longer accretion. Higher accretion rates imply stronger neutrino emission, stronger neutrino heating, and also generally a higher luminosity, which roughly balances neutrino heating under steady-state conditions in the pre-explosion phase \citep{murphy_12,mueller_15a}.

Similarly, several factors increase the GW emission if a progenitor evolves toward shock revival. Shock expansion increases the mass in the gain region 
and hence the neutrino heating and $L_\mathrm{turb}$. The Mach number also increases toward the transition to runaway shock expansion \citep{mueller_15a}.
Because of the critical influence of neutrino heating on the high-frequency GW emission, there may also be a dependence of the high-frequency signal on the EoS, with softer EoSs favoring stronger GW emission \citep{marek_08} aside
from providing stronger heating and generally more favorable conditions for explosion \citep{janka_12b,suwa_13,yasin_20}.

\subsection{Other Gravitational Wave Signal Features}
Other components of the supernova GW signal are not expected to be universal. These will, however, often hold particularly interesting clues about supernova dynamics, the progenitor structure, or the physics of matter at high densities.

\begin{figure}
    \centering
    \includegraphics[width=\linewidth]{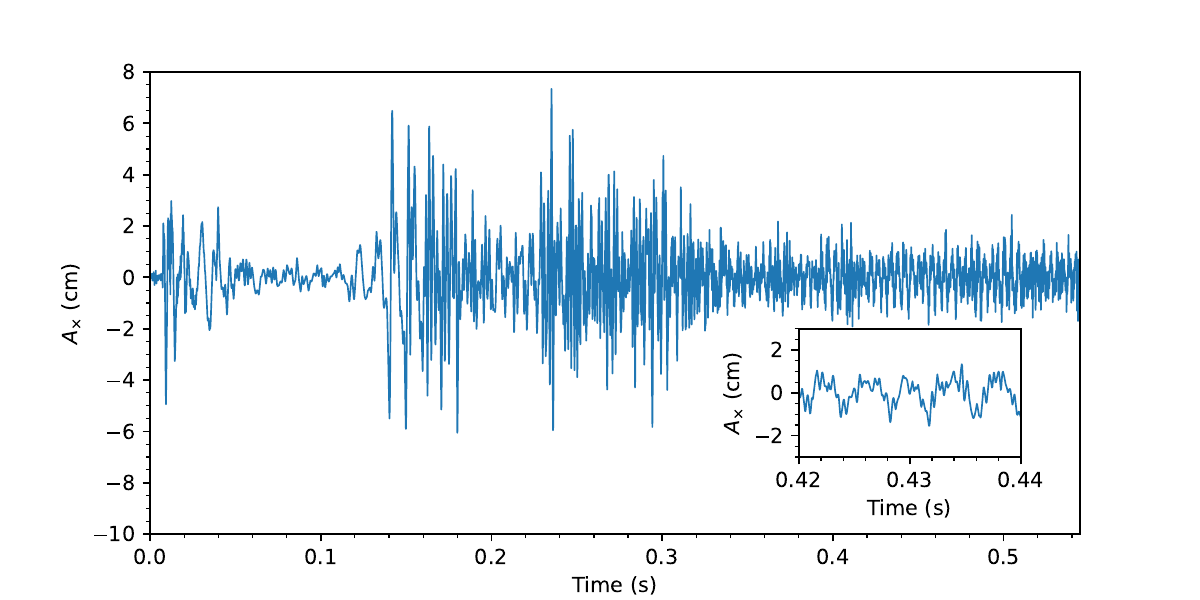}
    \caption{Predicted GW amplitude for a non-exploding, SASI-dominated supernova model 
    of an $18\msun$ star for one observer direction. Note the distinctly different periodicity from the high-frequency signal in Figure~\ref{fig:gw_template} that can be seen in the inset.}
    \label{fig:sasi_ampl}
\end{figure}

\begin{figure}
    \centering
    \includegraphics[width=\linewidth]{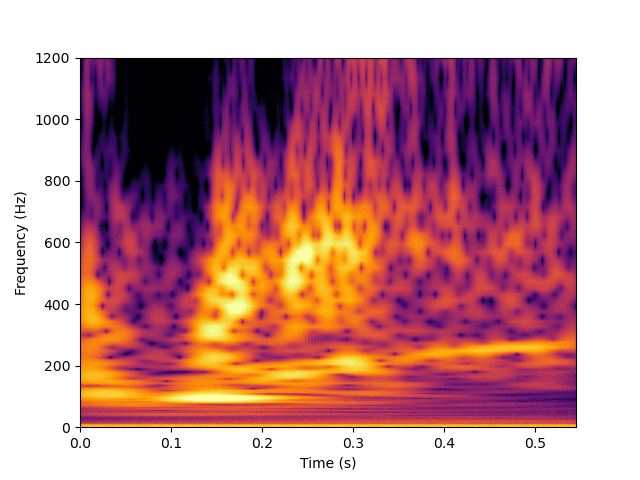}
    \caption{Spectrogram corresponding to the waveform of the SASI-dominated $18\msun$ model in Figure~\ref{fig:sasi_ampl}. The high-frequency signal is present from about $0.1\,\mathrm{s}$ to $0.3\,\mathrm{s}$ after bounce. In addition there is a low-frequency signal that rises from $100\texttt{-}200\,\mathrm{Hz}$ initially to more than $200\,\mathrm{Hz}$ after $0.4\,\mathrm{s}$.}
    \label{fig:sasi_spec}
\end{figure}

\subsubsection{The SASI signal}
The SASI contributes a GW signal at distinctly lower frequencies, typically of order $100\texttt{-}200\,\mathrm{Hz}$ (Figures~\ref{fig:sasi_ampl} and \ref{fig:sasi_spec}). While generally weaker than the high-frequency signal, it falls in the optimal sensitivity range for the currently operating LIGO, VIRGO and KAGRA detectors. The SASI signal tends to appear more consistently in non-exploding models \citep{kuroda_16b,andresen_17,radice_19}, but has also been reported in some models with small shock radii before explosion \citep{andresen_17}. It is quite commonly seen in models of very high mass progenitors \citep{shibagaki_21,powell_21} that eventually form black holes.
In this case, the signal frequency may ramp up to several hundred Hz as the proto-neutron star and the shock contract.
Certain nuclear equations of state (EoS) like the SFHx EoS \citep{steiner_10} and the CMF EoS \citep{motornenko_20} have been found to favor SASI activity \citep{kuroda_16b,powell_21,powell_26}.

The SASI frequency potentially holds information about the shock trajectory as it is determined by the advection and sound-crossing time between the shock and the proto-neutron star surface. Empirically, the SASI
period $T_\mathrm{SASI}$ is related to the shock radius $r_\mathrm{sh}$ and the proto-neutron star radius $R$ as
\begin{equation}
    T_\mathrm{SASI}
    \approx
    19\,\mathrm{ms}
    \times \left(\frac{r_\mathrm{sh}}{100\,\mathrm{km}}\right)^{3/2}
\times \ln\left(\frac{r_\mathrm{sh}}{R}\right),
\end{equation}
which can be motivated from the scaling of the advection time scale between the shock and proto-neutron star surface \citep{mueller_14}.
However, the relation between the SASI mode frequency and the SASI frequency in predicted spectrograms is not trivial. The SASI feature is often not clearly defined in frequency due a number of complications, e.g., intermittent SASI activity and potential frequency doubling \citep{andresen_17}. Despite these difficulties, the SASI feature has the potential to provide qualitative information about the dynamics of collapse and explosion. For example, the frequency trend and duration of the SASI signal can be used to distinguish successful explosion, black-hole formation without shock revival, and black-hole formation in stars with successful shock revival \citep{powell_25}.

\subsubsection{The core $g$-mode and other proto-neutron star oscillation modes}
\label{sec:core_g}
The prominent ramp-up signal is not the only oscillation mode reflected in predicted GW spectrograms. Additional $g$-modes and $p$- modes are often present \citep{cerda_13,torres_18}, though usually at low amplitude, and often merged into a forest of broad-band noise. In particular, there is often a $p$-mode forest above the ramp-up signal prior to and around the onset of shock revival.

Features attributable to specific modes are of particular interest because they could provide complementary information about the proto-neutron star structure. In particular, a \emph{core} $g$-mode sometimes leaves discernible traces in the spectrogram. This mode sometimes shows up as a gap
in the spectrogram associated with an avoided crossing with the $f$-mode or $p$-modes.
\citep{morozova_18,sotani_20,vartanyan_23}.
In some cases, the mode appears directly as an extra emission band in the spectrogram \citep{jakobus_23,sykes_26}. This core $g$-mode mostly involves oscillations of the convectively stable region between the low-entropy proto-neutron star core and the proto-neutron star convection zone. Since the mode lives at high density, its frequency serves as a probe for the properties of nuclear matter above saturation density.  Noting that the entropy gradient between the core and the proto-neutron star convection zone and its location in mass coordinate are not very progenitor-dependent, one can work out that
the mode frequency is primarily regulated by the lapse function, the thermodynamic derivative of pressure with respect to entropy, and the sound speed \citep{jakobus_25}
\begin{equation}
f_\mathrm{core}\propto \alpha^{5/2}c_s^{-1}
\left(\frac{\partial P}{\partial s}\right)^{1/2}_{\rho,Y_\mathrm{e}},
\end{equation}
though detailed model calculations are required for a precise prediction. When the core $g$-mode appears is still rather unclear. It has been found in 2D simulations with the CMF EoS for high-mass progenitors \citep{jakobus_23}, but has recently been reported in some 3D magnetohydrodynamic simulations of lower-mass stars as well
\citep{sykes_26}.

\subsubsection{Phase-transition signals}
\label{sec:gw_pt}
It is noteworthy that the appearance of the core $g$-mode may provide \emph{indirect} evidence of a phase transition at several times the nuclear saturation density, since such a phase transition requires the EoS to be relatively at lower densities to accommodate current constraints on the maximum neutron star mass. A \emph{first-order} phase transition could produce a more direct ``smoking gun'' in the GW signal.
As shown by Zha et al.\ \citep{zha_20}, the second collapse and bounce in the phase-transition scenario for supernova explosions can generate a powerful short burst signal
with strain amplitudes of several $10^{-21}$ at $10\,\mathrm{kpc}$. These predictions have, however, been obtained under idealized conditions with an EoS not compatible with current experimental and astrophysical constraints. Replication in 3D simulations with a more realistic EoS (for which the second collapse would happen much later than in \citep{zha_20}) are still pending. Another challenge for the detection of such a phase transition consists in the high-frequency of the signal with most of the power at several kHz.

\subsubsection{Post-bounce signatures of rotation and magnetic fields}
Other stable or unstable oscillatory modes can develop during the post-bounce phase in the presence of rotation and magnetic fields. For sufficiently rapid rotation, triaxial instabilities driven by a corotation resonance can occur already during or shortly after bounce
\citep{ott_07b,scheidegger_10}. They can also develop later and give rise to strong and long-lasting GW emission in the range of several hundred Hz \citep{kuroda_19,shibagaki_20}. For a typical Galactic supernova at $\mathord{\sim}10\,\mathrm{kpc}$, strain amplitudes of $\mathord{\sim}10^{-20}$ can be attained \citep{shibagaki_20}.
The high amplitudes and rather long coherence times imply a large detection horizon up to Mpc scales with third-generation instruments. This is balanced by the lower expected rates. The requisite rapid rotation is only expected in hypernovae, i.e., in about $1\%$ of the supernova population \citep{smith_11}.

Prior to shock revival and during the early explosion phase, strong magnetic fields affect GW emission via their impact on the explosion dynamics, but their qualitative impact on the GW signal structure is more limited \citep{jardine_22,powell_23}. In particular, magnetorotational explosion tend to be characterized by high amplitudes and a strong tail signal form anisotropic shock expansion. Statistical classification techniques may still be able to discriminate between neutrino-driven and magnetorotational explosions based on the quantitative changes in the spectrogram and the absence of presence of a rotational bounce signal \citep{powell_24}.
Moreover, long-term dynamo field amplification in the proto-neutron star over time scales of seconds may produce a genuine signature of magnetic fields in GWs in the form of low-frequency excess power at late times and distinct spikes from inertial modes \citep{raynaud_22}.

\subsection{Neutrino Memory and Asymmetric Shock Expansion}
In addition to the memory signal
from anisotropic
neutrino emission \citep{epstein_78}, a very-low-frequency contribution to the supernova GW signal comes from anisotropic shock expansion
during the explosion phase. The neutrino memory
(Equation~\ref{eq:gw_nu}) is determined by the overall degree of emission anisotropy.
The direction of emission anisotropy is not stable during the pre-explosion phase, so the build-up of the memory signal happens primarily after the onset of the explosion.
Even during this phase, the direction of anisotropy is not completely stable and may exhibit modulations on time scales of several hundred milliseconds to seconds \citep{choi_24,powell_24b}. Distance-normalized strain amplitudes as large as 
$h D\sim 1000\,\mathrm{cm}$ can be reached on time scales
of seconds in modern 3D simulations, though values of
several $100\,\mathrm{cm}$ are more common.

The matter memory effect was first described in 2D
simulations \citep{murphy_09}, where the deformation of the shock during the explosion phase is enhanced and stabilized by the presence of the grid symmetry axis.
The matter memory signal can be related to the monopole and quadrupole coefficients
$a_0$ and $a_2$ of the angle-dependent shock radius, the
compression ratio $\beta$ and adiabatic index $\gamma$
of the infalling matter. In axisymmetry, the
amplitude $A^{\mathrm{E}2}_{20}$ is roughly
\begin{equation}
A^{\mathrm{E}2}_{20}
=
\frac{256\pi^{3/2}}{5\sqrt{15}}
\bar\rho_\mathrm{pre}
(\beta-1)
a_0^3
\left[(4+\gamma)a_2\dot a_0
+\dot a_2 a_0\right],
\end{equation}
where $\bar\rho_\mathrm{pre}$ is the angle-averaged pre-shock density. In 3D simulations, the tail signal from asymmetric shock expansion is typically weaker than in 2D models, with $hD$ reaching values of a few cm \citep{choi_24} to tens of cm in magnetorotational models \citep{powell_23,powell_24b}. Thus, the matter memory is generally subdominant to the neutrino memory.

The memory signal is the dominant contribution to the overall GW spectrum at frequencies of a few tens of Hz
and lower, and its predicted nominal signal-to-noise ratios is higher than for the other signal components \citep{choi_24}. Due to its low-frequency nature, the memory signal is a particularly interesting target for future lunar-based \citep{jani_21,ajith_25} or space-based detectors in the deci-Hz range \citep{harry_06,yagi_11,kawamura_21}. However, the non-oscillatory nature of the memory requires careful consideration about signal extrapolation, windowing and the nature of detector noise to reliably assess its detectability.

\section{Conclusions -- Towards the Detection of a Galactic Supernovae}
\label{sec:conclusions}
Modern supernova simulations thus predict a broad range of signal features that can reveal the inner workings of exploding massive stars qualitatively and quantitatively. Some of these features are expected to be generic, while others would reveal unknown physics or exotic progenitor and explosion scenarios.

In the event of a Galactic supernova, the challenge will be to actually detect these features and extract as much information from them as possible. Much progress has been made on this data analysis problem as well, both with regard to the classification of signal features and different supernova explosion scenarios \citep{logue_12,roma_19,powell_24,richardson_24} and with regard to quantitative parameter estimation \citep{astone_18,bizouard_21,powell_22,afle_23,murphy_24,sasaoka_24}. 
The mere detection of explosions will  be a smaller challenge.
With current instruments, a Galactic event will likely be observable
out to distances of a few kpc, with larger detection horizons for very energetic explosions
\citep{powell_19,powell_20,mezzacappa_23,powell_23,szczepanczyk_24,choi_24}.
Detection ranges will increase by about an order of magnitude with third-generation instruments, typically encompassing the entire Milky Way and its satellites. For magnetorotational explosions, detections beyond $1\,\mathrm{Mpc}$ may become possible \citep{powell_23}. The neutrino memory signal may also be detectable to such large distances both in ground-based
and space-based deci-Hertz detectors \citep{choi_24}, though the detection of a memory signal involves subtleties due to its non-oscillatory character \citep{richardson_22}.
For quantitative parameter inference or classification of different explosion mechanisms, the supernova typically needs to occur closer than the horizon distance by a factor of a few. In this case, it becomes possible
to constrain, e.g., the peak frequency, duration and slope of the high-frequency signal, or the frequency of the SASI signal
\citep{powell_22,murphy_24,powell_25,sierra_26}. Hence, such measurements will be comfortably feasible with third-generation instruments for most Galactic events, though a very nearby supernova would be required with current instruments. In the case of rapid rotation, a measurement of the core angular momentum would already be possible at distances of several kpc with current detectors \citep{abdikamalov_14}.
We refer to the contribution by M.~Zanolin
in this volume for details on the problem of signal detection and analysis.

This progress should not obscure the formidable effort that is required to prepare for a Galactic supernova.
Such an event will arguably be the biggest breakthrough in multi-messenger astronomy as yet, but the very combination of many messengers and potentially ``messy'' signals prone to model-dependent interpretation requires a transformative approach to maximize the potential for scientific discovery.

Insufficient consideration of systematic errors and an omission or eclectic interpretation of complementary information from neutrinos, gravitational waves, and electromagnetic radiation may result in a proliferation of biased and mutually contradictory conclusions instead of truly constraining supernova physics. Robust conclusions from
multi-messenger observations will require proper uncertainty quantification by means of model ensembles and multi-model comparisons, as well as a consensus process for qualitative assessment of uncertainties and quality of evidence as familiar, e.g., from the IPCC (Intergovernmental Panel on Climate Change) reports \citep{mastrandrea_11}. 
Strong integration between theory and observations across the different multi-messenger domains on an equal footing is required to assess evidence, develop strategies for hypothesis testing and parameter inference and to identify knowledge gaps. The theory ecosystem also needs to adapt (and be supported) to deliver input for this purpose. This entails, e.g., the development of integrated and increasingly automated modeling pipelines and curated databases with a well-maintained complementary simulation tools. Recognizing the upcoming technical and organizational challenges in explosion modeling will be key to the success of supernova gravitational wave astronomy.

\section*{Acknowledgments}
I extend my thanks to the organizers of the
First IGWN Symposium on Core Collapse Supernova Gravitational Wave Theory and Detection, and to the other participants for fruitful discussions at this symposium. I acknowledge support from the ARC through Discovery Projects DP240101786 and DP260104967, as well as computer time allocations from Astronomy Australia Limited's ASTAC scheme, the National Computational Merit Allocation Scheme (NCMAS). Some of this work was performed on the Gadi supercomputer with the assistance of resources and services from the National Computational Infrastructure (NCI), which is supported by the Australian Government, and through support by an Australasian Leadership Computing Grant.  

\section*{Data Availability}
The data from our simulations will be made available upon reasonable requests made to the author.

\bibliography{paper}

\end{document}